\documentclass[12pt]{iopart}

\usepackage{graphicx}
\usepackage{subfigure}
\usepackage[top= 1.8 cm, bottom=1.8 cm, left=2.4 cm, right= 2.4 cm]{geometry}

\begin{document}

\title[]{On loading of a magneto-optical trap on an atom-chip with U-wire quadrupole field}

\author{Vivek Singh\textsuperscript{1}, V. B. Tiwari\textsuperscript{1,3}, K. A. P. Singh\textsuperscript{2}, S. R. Mishra\textsuperscript{1,3}}

\address{\textsuperscript{1}Laser Physics Applications Section,
                             Raja Ramanna Center for Advanced Technology, Indore-452013, India.}
\address{\textsuperscript{2}Pulsed High Power Microwave Division, Raja Ramanna Center for Advanced Technology, Indore-452013, India.}
\address{\textsuperscript{3}Homi Bhabha National Institute, Mumbai-400094, India.}

\ead{viveksingh@rrcat.gov.in}
\vspace{10pt}
\begin{indented}
\item[]December 2017
\end{indented}

\begin{abstract}
The role of U-wire generated quadrupole like magnetic field configuration on loading of cold $^{87}Rb$ atoms in a magneto-optical trap (MOT) on atom-chip has been investigated. It is shown by simulations that the quadrupole field profile and the center of the quadrupole field vary depending upon the U-wire current and bias field values. The result of simulations show that the resemblance of generated field configuration to the ideal quadrupole field critically depends on the values of current in U-wire and bias magnetic field. The observed number of trapped atoms in the MOT and its position has also been found to depend on the quadrupole field configuration. The number of trapped atoms in the MOT reaches maximum when the field configuration approaches close to an ideal quadrupole field. This study will help in optimising the number in the MOT before transfer of these cold atoms in the atom-chip wire trap.
\end{abstract}

%
%
%
%
%
\section{Introduction}
Atom-chip \cite{Folman1, Folman2}, with micro-structured wires on a reflecting surface, provides a versatile platform for trapping, guiding and splitting of cold atom cloud on a miniaturized scale. It is expected to play the role of the basic building block in atom-optic devices such as precision inertial sensors \cite{Dick}, gravimeters \cite{abend}, matter-wave interferometers \cite{schumm} and precision sensing of DC magnetic field \cite{wilder} as well as microwave field \cite{Bohi} with high spatial resolution. The cold atoms used in these atom optic devices are commonly produced by a magneto-optical trap (MOT) \cite{raab} in the temperature range of few micro-Kelvin to few hundreds of micro-Kelvin. To load a MOT with cold atoms, the atomic source could be the background vapor in the chamber or an atomic beam slowed down suitably by a Zeemam slower. The standard MOT formation involves six beams intersecting at the center of quadrupole field generated from a pair of anti-Helmholtz coils. However, this six beam geometry is not suitable for a MOT to be formed near the atom-chip surface. This requirement can be fulfilled by using a mirror-MOT configuration \cite{clifford} which is commonly used to load the atoms near the atom-chip surface. In the mirror-MOT near the atom-chip surface, the quadrupole field can be generated by either a pair of anti-Helmholtz coils or by using a current carrying U-shaped wire behind the atom-chip and along with bias fields. The experimental set-up in the former case becomes more complicated with little optical access available near the central region for the MOT. The latter scheme involving U-wire for generation of quadrupole like magnetic field, called U-MOT scheme, provides better optical access, efficient transfer of cold atoms from U-MOT to micro-trap of atom-chip wire \cite{wilder2} and makes the set-up more compact. This can be helpful to produce the Bose-Einstein condensation (BEC) faster \cite{aubin, hansel} in time with less stringent requirement of vacuum. However, the studies on the loading of this U-MOT and its optimisation are limited and not reported in detail in the literature. \\
 Here, we report our studies on the loading of U-MOT for $^{87}Rb$ atoms on the surface of an atom-chip with U-wire generated quadrupole field. We have first simulated the magnetic field generated due to current carrying U-wire and bias fields. There are two homogeneous bias fields applied in the set-up, parallel ($B_{y}$) and perpendicular ($B_{z}$) to the plane of the U-wire. The effect of bias fields (for a given current in U-wire) on the structure of the quadrupole field and its center has been investigated. The results of simulations show that generated field configuration is close to ideal quadrupole field configuration for a set of parameters including current in U-wire and bias magnetic field values. Experimentally, we have observed that the number of cold atoms in the U-MOT is maximum for those values of current and bias field which result in the magnetic field configuration to approach closer to the ideal quadrupole field (shown by simulations) with appropriate field gradient. We have found that in our U-MOT set-up, the optimum values of bias magnetic fields are $B_{y}$ $ \sim $ 8.0 G and $B_{z}$ $ \sim 3.5 G$ for a U-wire current of 40 A, which result in the maximum number of cold atoms $\sim 6.6 \times 10^{6}$ for cooling laser beam intensity $\sim$ $8.5 mW/cm^{2}$ and detuning of $-14 MHz$. These studies are, in general, useful in optimising the number of atoms in the U-MOT before transfer of these cold atoms to micro traps of an atom-chip.
\section{Simulation of quadrupole field for U-MOT}

\begin{figure}[ht]
  \centering
 \includegraphics[width=8.0 cm]{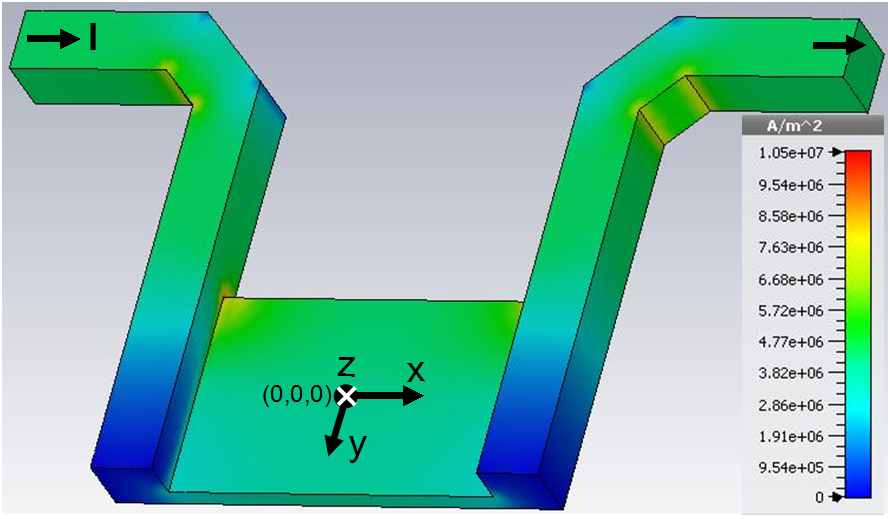}
 \caption{ Schematic 3D-view of U-wire. The central broad portion of U-wire has dimensions 1 mm (thickness) $\times$ 10 mm (width) $\times$ 15 mm (length). The side legs have dimensions 3 mm $\times$ 3 mm $\times$ 22 mm. The central part is 2 mm down from the top surface of the side legs. The simulated current density distribution in the U-wire is shown in color code. The red (blue) colors correspond to high (low) current densities. }
  \end{figure}
Working of a U-MOT depends on the correct orientation of quadrupole field with respect to polarisation of laser light beam in each direction. Therefore, it is essential to calculate the field configuration of the magnetic field due to U-wire and bias field. The simulations have been performed for the U-wire whose schematic is shown in fig 1. We have first simulated the complex current density in the U-wire using CST (Computer Simulation Tool) software. The value of U-wire current was kept at 40 A as per experimental limitation on the current due to heating of chip with higher value of current. It is clear from the fig 1 that current density is nearly homogeneous in central part of the U-wire. The current densities are higher on the edges. After having information about current density distribution in the U-wire, we have simulated the magnetic field for different values of bias field. The quadrupole like field could be formed by a appropriate combination of field due to U-wire and uniform bias field.  \\

 \begin{figure}[ht]
      \centering
      \subfigure[]{
       \resizebox*{7.0cm}{!}{\includegraphics{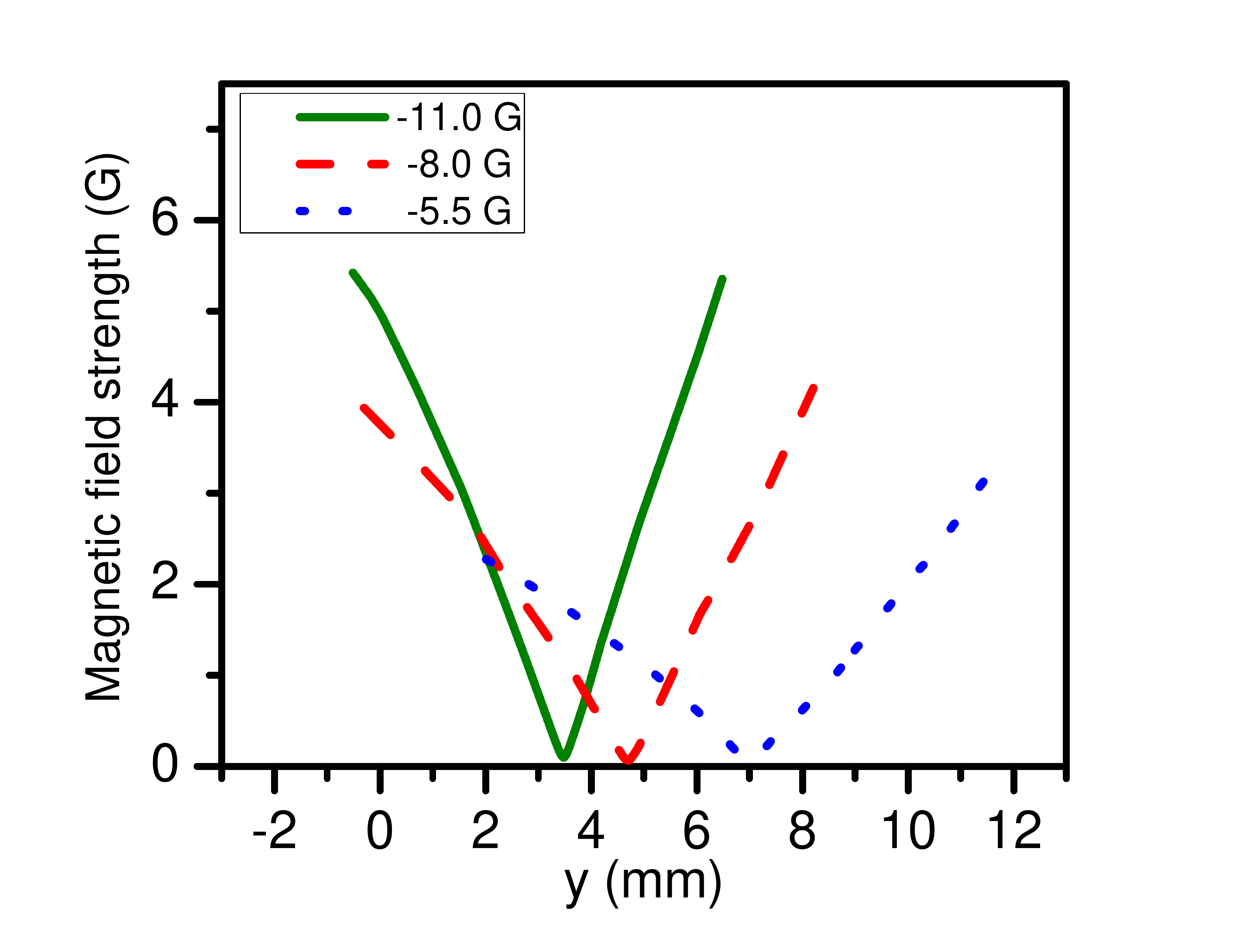}}}\hspace{0pt}
       \subfigure[]{
       \resizebox*{6.6cm}{!}{\includegraphics{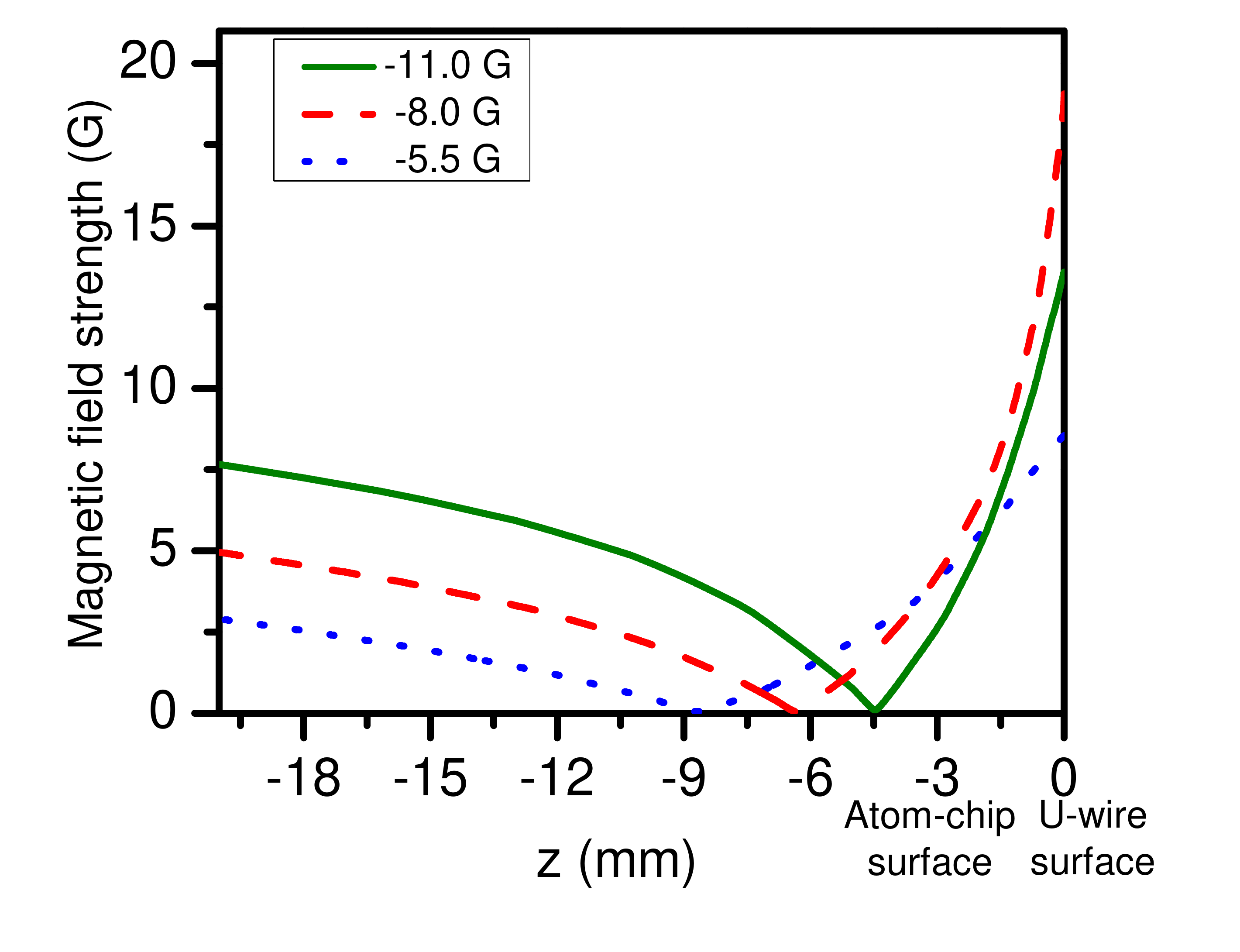}}}
       \caption{ Calculated variation in magnetic field amplitude quadrupole like field in y-direction (a) and z-direction (b) at 40 A of current in U-wire for different values of bias fields in y-direction ($B_{y} \neq 0, B_{z} = 0$). } \label{sample-figure}
       \end{figure} 
 Fig 2(a) and 2(b) show the calculated variation of field magnitude along y- and z-direction respectively for different bias fields. The quadrupole field gradient vector near the quadrupole center was found to be (5, 17, 19) G/cm for bias field of 11.0 G, (3.5, 11, 12) G/cm for bias field of 8.0 G and (2, 5, 4.5) G/cm for bias field of 5.5 G. It is to be noted that the position of the quadrupole field center is shifted along y- and z-directions for different bias field values. This shift is an important parameter to asses the position of the MOT center from the atom-chip surface and the laser beams may be aligned accordingly for MOT formation.\\
 \begin{figure}[ht]
  \centering
  \subfigure[]{
   \resizebox*{4.6cm}{!}{\includegraphics{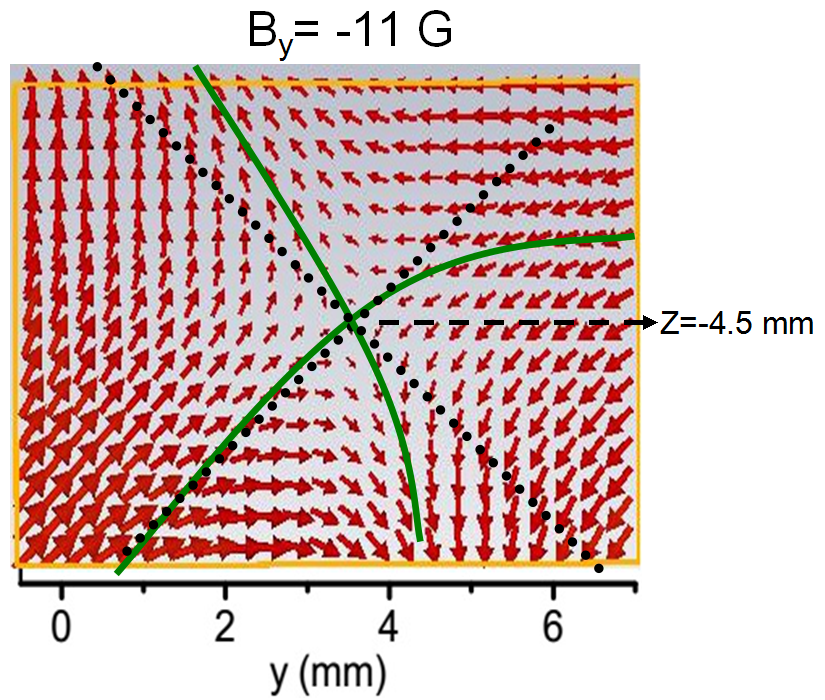}}}\hspace{0pt}
   \subfigure[]{
   \resizebox*{4.6cm}{!}{\includegraphics{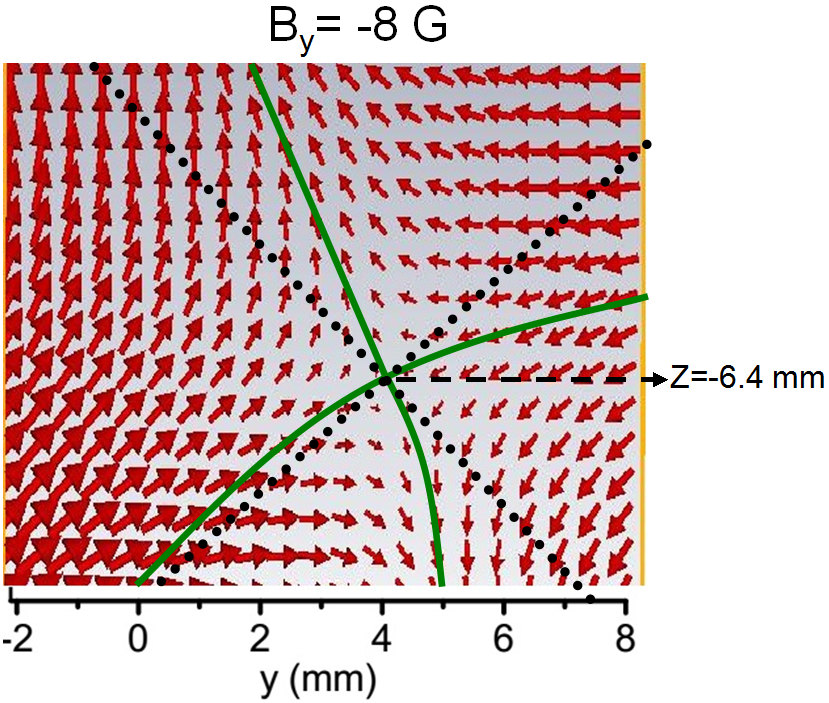}}}\hspace{0pt}
   \subfigure[]{
     \resizebox*{4.4cm}{!}{\includegraphics{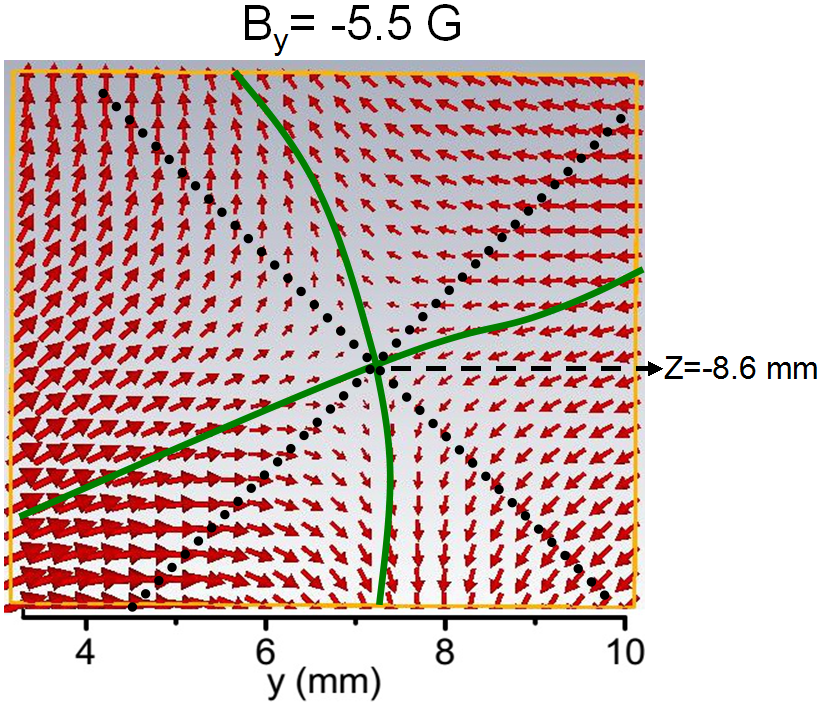}}}
   \caption{ Vector field plots of resulting quadrupole field in y-z plane (x=0) for 40 A current in U-wire and different values of bias field $B_{y}$. (a) for $B_{y} = -11$ G (b) for $B_{y} = -8 $ G (c) for $B_{y} = -5.5$ G. The dashed line represents the axes of ideal quadrupole fields and the continuous curve represents the corresponding axes of simulated field.} \label{sample-figure}
   \end{figure}
 Figure 3 shows the vector field plots of quadrupole field for $B_{z}=0$ and different values of bias field ($B_{y}$). As, the bias field was reduced from 11.0 G to 5.5 G, the vector field lines were deviated more from ideal quadrupole axes (dotted lines). This is due to the shifting of minimum position in y-direction with reduction in the $B_{y}$. The minimum position in y-direction is shifted by $\sim 4 mm$ as $B_{y}$ is reduced from 11.0 G to 5.5 G, as shown in figs 2 and 3. Ideally, one would want the minimum at $y=0$, which would require much higher value of $B_{y}$, leading to much steeper field slope and reduced MOT volume. This is the probable cause for reduction in number of atoms in MOT as magnitude of bias field $B_{y}$ was increased from 8.0 G to 11.0 G, as discussed latter in section-3.\\
 
  \begin{figure}[]
                \centering
                \includegraphics[width=8 cm]{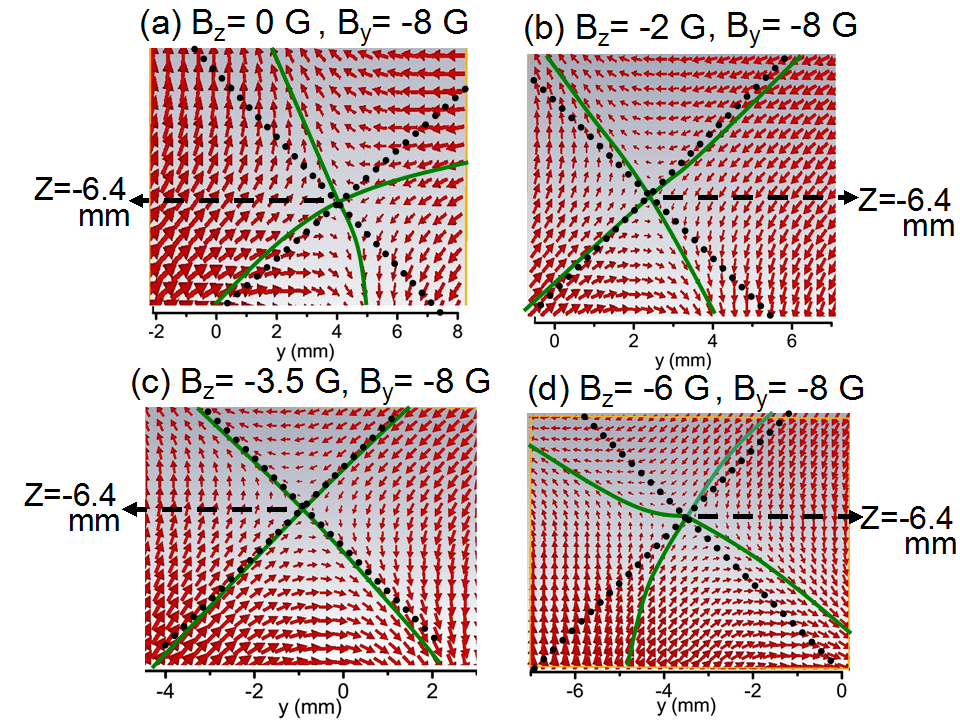}
                \caption{ Vector field plots of resulting quadrupole field in y-z plane (x=0) for 40 A current in U-wire. Here, graphs (a), (b), (c) and (d) are for $B_{y} = -8$ G and bias field $B_{z}$ = 0, -2, -3.5 and -6 G, respectively. The dashed line represents the axes of ideal quadrupole fields and the continuous curve represents the corresponding axes of simulated field.}
                \end{figure}
 Further, in order to bring the quadrupole field center near to $y=0$ (which is center of atom-chip and U-wire), we introduce the z-component ($B_{z}$) in the bias field by using two separate Helmholtz coils. Figure 4 shows simulated vector field plot for different values of $B_{z}$ for a given $B_{y}$ = - 8.0 G. As bias field in negative z-direction is increased, the quadrupole field center gets shifted towards y = 0 and the obtained quadrupole field axes (solid curve) are more aligned to the ideal quadrupole field axes (dashed line). For $B_{z}$ = - 3.5 G field, these axes overlap each other and the obtained quadrupole field is close to ideal quadrupole field (fig 4(c)). At further higher value of $B_{z}$ magnitude, the field lines starts to deteriorate from ideal quadrupole field as shown in fig 4(d). Here we note that by introducing $B_{z}$, the quadrupole center is not shifted along z-direction, as expected. These studies reveal that for a 40 A current in U-wire, bias fields ($B_{y}$ = - 8.0 G, $B_{z}$ = -3.5 G) are optimum parameters for our experiments.

\section{Experimental setup and results}\label{class}

Atom-chip was fabricated using silicon (Si) substrate of size 25 mm $\times$ 25 mm (thisckness $\sim 1 mm$). A layer of $SiO_{2}$ of thickness 50 nm was deposited on Si substrate to make insulating layer. In order to increase the adherence strength between the $SiO_{2}$ layer and the gold layer, a thin layer of $\sim 50$ nm thickness of Chromium (Cr) is coated on the substrate before depositing the gold (Au) of 2 $\mu$m. Two wires of U-shape (200 $\mu$m width) and one wire of Z-shape (200 $\mu$m width) are engraved on the gold coated surface of chip using photo-lithography technique as shown in fig. 5(a). This chip served as a mirror surface for U-MOT formation.\\
\begin{figure}
\centering
\subfigure[]{
\resizebox*{5.0 cm}{!}{\includegraphics{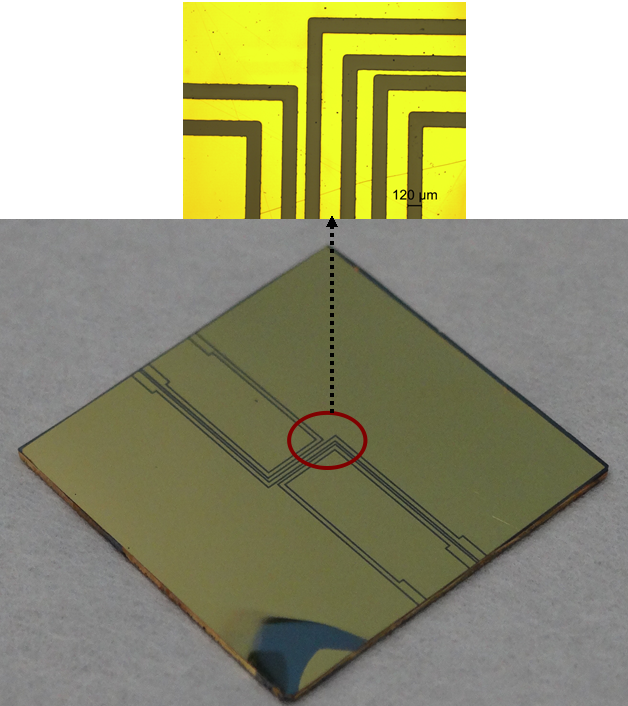}}}\hspace{0pt}
\subfigure[]{
\resizebox*{4.0 cm}{!}{\includegraphics{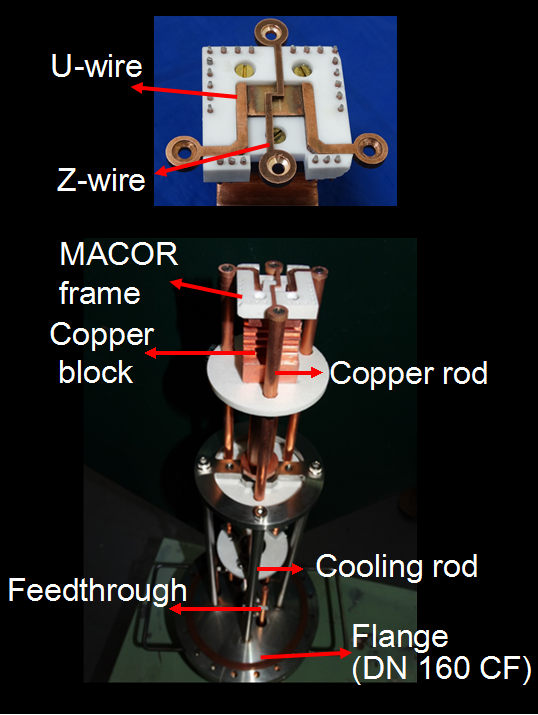}}}
\caption{(a) Photograph of atom chip shows the two U and one Z-shaped gold wire. The inset in the figure show parts of z- and u-shaped wires of atom chip. (b) Photograph of atom-chip mounting system showing U-wire fixed into a MACOR ceramics holder. The Z-wire seen here will be used for future studies.} \label{sample-figure}
\end{figure}  
This atom-chip is placed on top of a mount system made on a DN 160 CF flange. This flange has four SS rods welded on it, on which a MACOR frame is fixed to hold the copper U-wire, copper Z-wire and atom-chip (on the top). The MACOR frame has appropriate grooves for these wires and atom-chip. The U-wire is located at $\sim$ 3 mm separation from the atom-chip. The MACOR frame is also connected to a solid copper block (which is connected to SS rods) for the heat removal as shown in fig.5. There are four high current copper (Cu) feed-throughs ($\sim$200 A capacity) and a 28-pin low current(few amperes on for each pin) feed-through fixed on the atom-chip mount flange as shown in fig. 5(b). These high current feed-throughs can be connected to U-wire and Z-wire ( not used in the present work) to create the magnetic fields needed for the MOT formation and initial magnetic trapping.\\ 
The experimental setup has an octagonal shaped vacuum chamber. This octagonal chamber is connected to a six-way cross on top of it using a $\sim$ 100 mm long I-piece. The atom-chip mount system is placed inside the octagonal chamber through this six-way cross such that chip surface is positioned $\sim$ 7 mm above the center of the octagonal chamber. The vacuum pumps (turbo molecular pump (TMP) (77 l/s) and sputter ion pump (SIP) (300 l/s)) are connected to this six-way cross for evacuating the chamber upto $\sim 6\times10^{-9}$ Torr. During the U-MOT operation, a current of 40 A is supplied to U-wire in presence of variable bias fields $B_{y}$ and $B_{z}$. The central part of this U-wire has been kept broad ($\sim$ 10 mm) to get a quadrupole field profile more close to the ideal one \cite{wilder2, pkruger}.  \\
\begin{figure}[ht]
             \centering
              \includegraphics[width=6.7 cm]{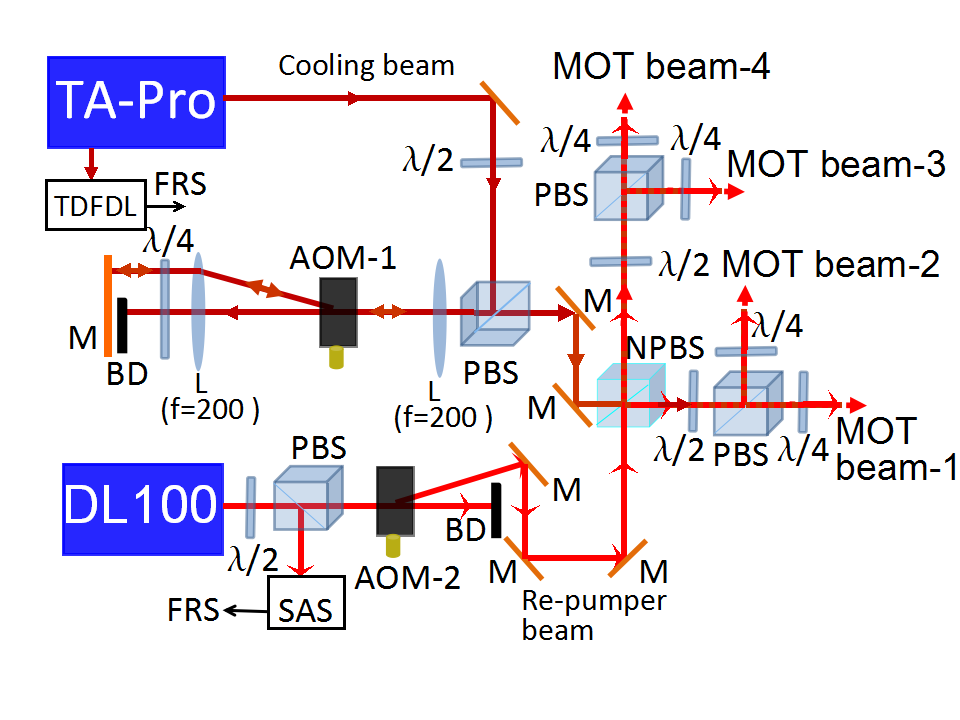}
              \caption{ Schematic drawing of the light manipulation area. The laser beams of toptica photonics laser systems (TA-Pro and DL100) are split into four beams. Laser light from the TA-Pro and the DL-100 and their frequencies are shifted by AOMs. BD; beam dump, L; Lens, TDFDL; tunable Doppler free dichroic lock, SAS; saturated absorption spectroscopy, FRS; frequency reference signal.}
              \end{figure}
The optical layout used in the setup is shown schematically Fig. 6. We have used a commercial oscillator-amplifier system (Toptica TA-Pro), capable of giving output power of $\sim$ 1 W at 780 nm, as a cooling laser to drive $F=2\rightarrow F^{'}=3$ transition for the $^{87}Rb$ MOT.  A second laser, (Toptica DL-100), capable of generating an output power of 70 mW, is used  as re-pumper laser to drive $F=1\rightarrow F^{'}=2$ transition of  $^{87}Rb$ atom. The cooling and re-pumper laser beams are passed through acousto-optical modulator (AOM-1 and AOM-2) in double pass and single pass configurations respectively, as shown in fig. 6. These two laser beams are mixed together using a non-polarizing beam splitter cube (NPBS). Four independent MOT beams are derived from this combined beam using suitable optics. The beams were expanded and truncated to the size of $\sim$ 17 mm diameter. Out of these four beams, two MOT beams (MOT beam-1 and MOT beam-2) with circular polarizations are incident on the atom-chip surface at $45^{0}$ angle. The other MOT beams(MOT beam-3 and MOT beam-4) with circular polarisations are aligned in counter propagating arrangement close to the chip surface (parallel to the surface) such that central part of the U-wire is along the propagation directions of the beams as shown in fig 7(a).\\
\begin{figure}[ht]
\centering
\subfigure[]{
\resizebox*{6.0 cm}{!}{\includegraphics{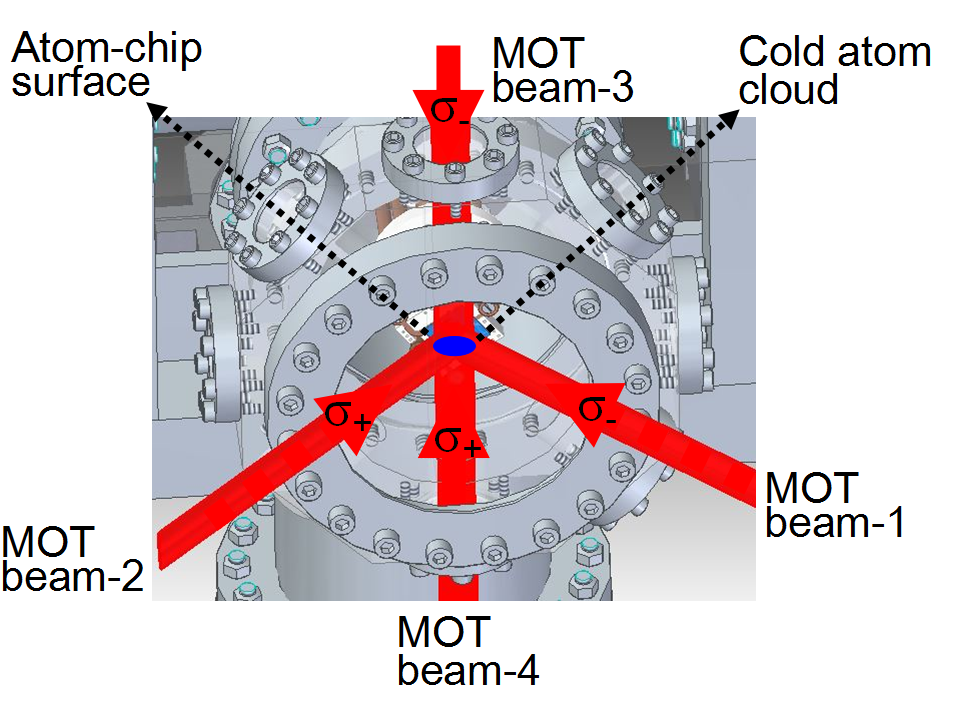}}}\hspace{0pt}
\subfigure[]{
\resizebox*{7.5 cm}{!}{\includegraphics{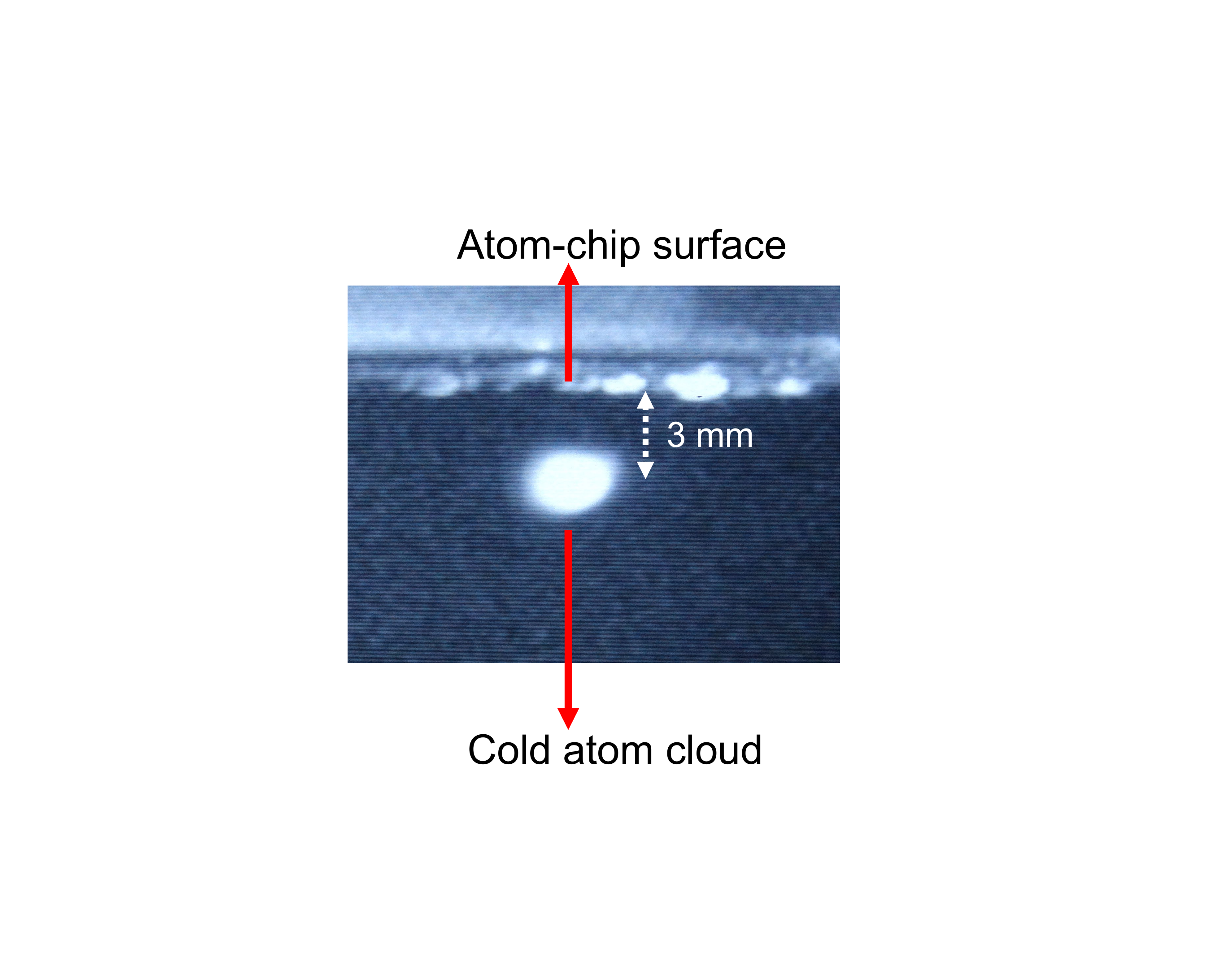}}}
\caption{(a) The schematic view of four MOT beams transmitted to the octagonal chamber for the atom-chip surface U-MOT. (b) The CCD camera image of cold $^{87}Rb$ atom cloud in the U-MOT near the atom-chip surface.} \label{sample-figure}
\end{figure}
Two pairs of coils in Helmholtz configuration are placed around the octagonal MOT-chamber to produce uniform bias magnetic filed in y- and z-directions. These Helmholtz coils along with U-wire are used to produce quadrupole like magnetic field required for the operation of U-MOT. A Rb-dispenser source to produce Rb vapour in the vacuum chamber is mounted on one of the view-port of the octagonal chamber.\\
We have used the tunable Doppler-free dichroic lock (TDFDL) technique developed in our lab to tune and stabilize frequency of the cooling laser \cite{vivek}. The quadrupole magnetic field was generated by flowing a 40 A current in U-wire and by applying $\sim$ 8.0 G bias magnetic field in negative y-direction. The quadrupole field strength was $\sim$ 3.5 G/cm along x-direction and $\sim$ 11-12 G/cm in radial direction (y and z directions) near the central region of the U-wire. The current in Rb-dispenser was maintained at 3.2 A during the experiments. The MOT cloud having cold $^{87}Rb $ atoms was formed at a distance of $\sim$ 3 mm below the atom chip surface ( as shown in fig. 7 (b) ). The fluorescence imaging technique using digital CCD camera and image processing software was used to estimate the number and temperature of the atom cloud in the MOT. The maximum number in the MOT was $\sim 3.7 \times 10^{6}$ at the temperature of $\sim$ 300 $\mu K$, for each cooling MOT beam intensity $\sim 8.5 mW/cm^{2}$ (detuning $\sim$ -14 MHz) and 40 A current in U-wire with bias fields $B_{y}$= - 8.0 G and $B_{z}$= 0.\\
The method of generation of quadrupole like field in the U-MOT is different from conventional MOT. In conventional MOT, the quadrupole field is normally generated using anti-Helmholtz coils. In the U-MOT, the quadrupole like field is generated by combination of U-wire with homogeneous bias magnetic filed. As we change the bias field for a fixed current in the U-wire, the quadrupole field center as well as the value of the field gradient change. Therefore, it is crucial to study the variation in the number of cold atoms in the U-MOT with different bias magnetic field to get optimum value of the bias field.\\ 
\begin{figure}[ht]
\centering
\subfigure[]{
\resizebox*{7.1 cm}{!}{\includegraphics{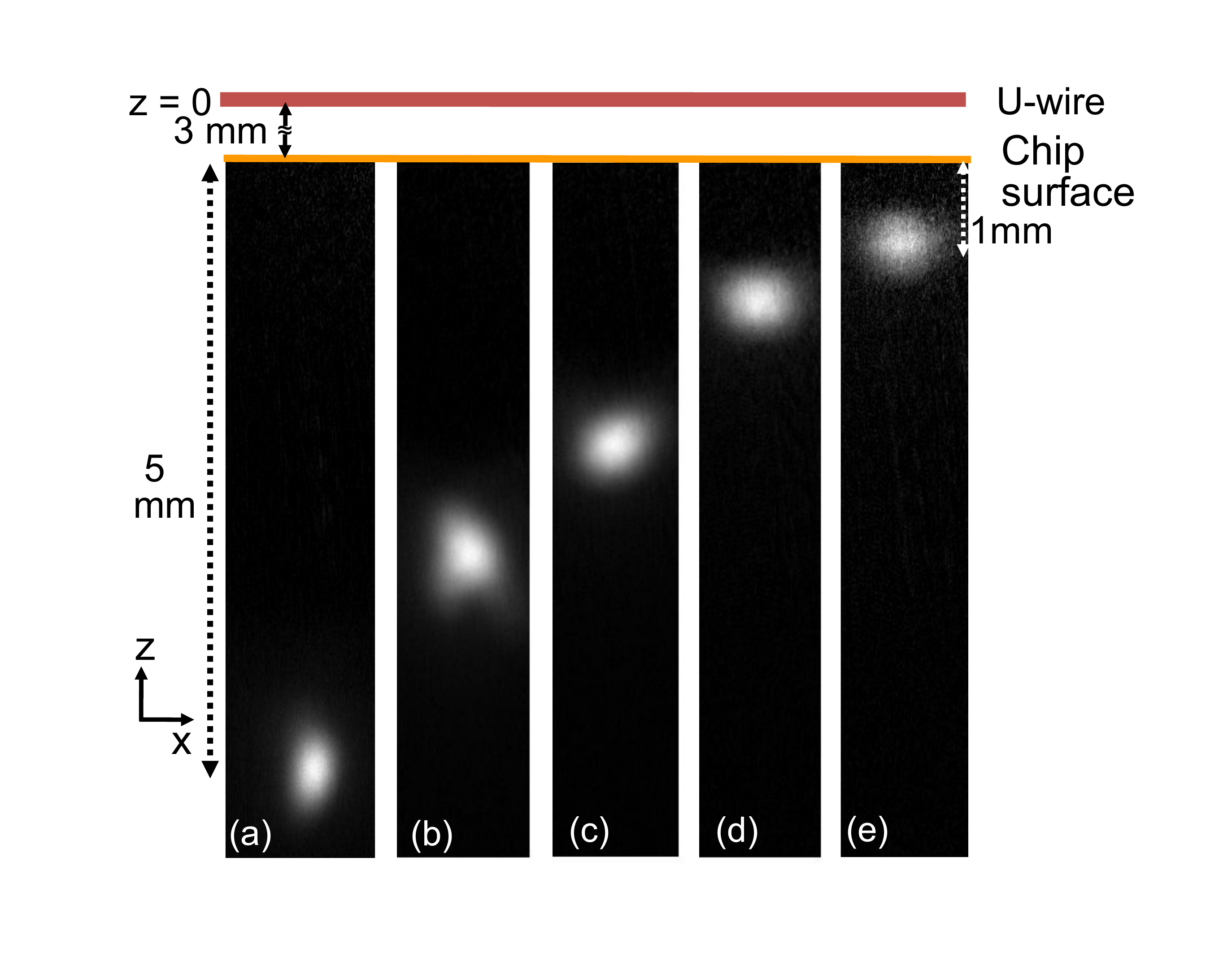}}}\hspace{0pt}
\subfigure[]{
\resizebox*{6.7 cm}{!}{\includegraphics{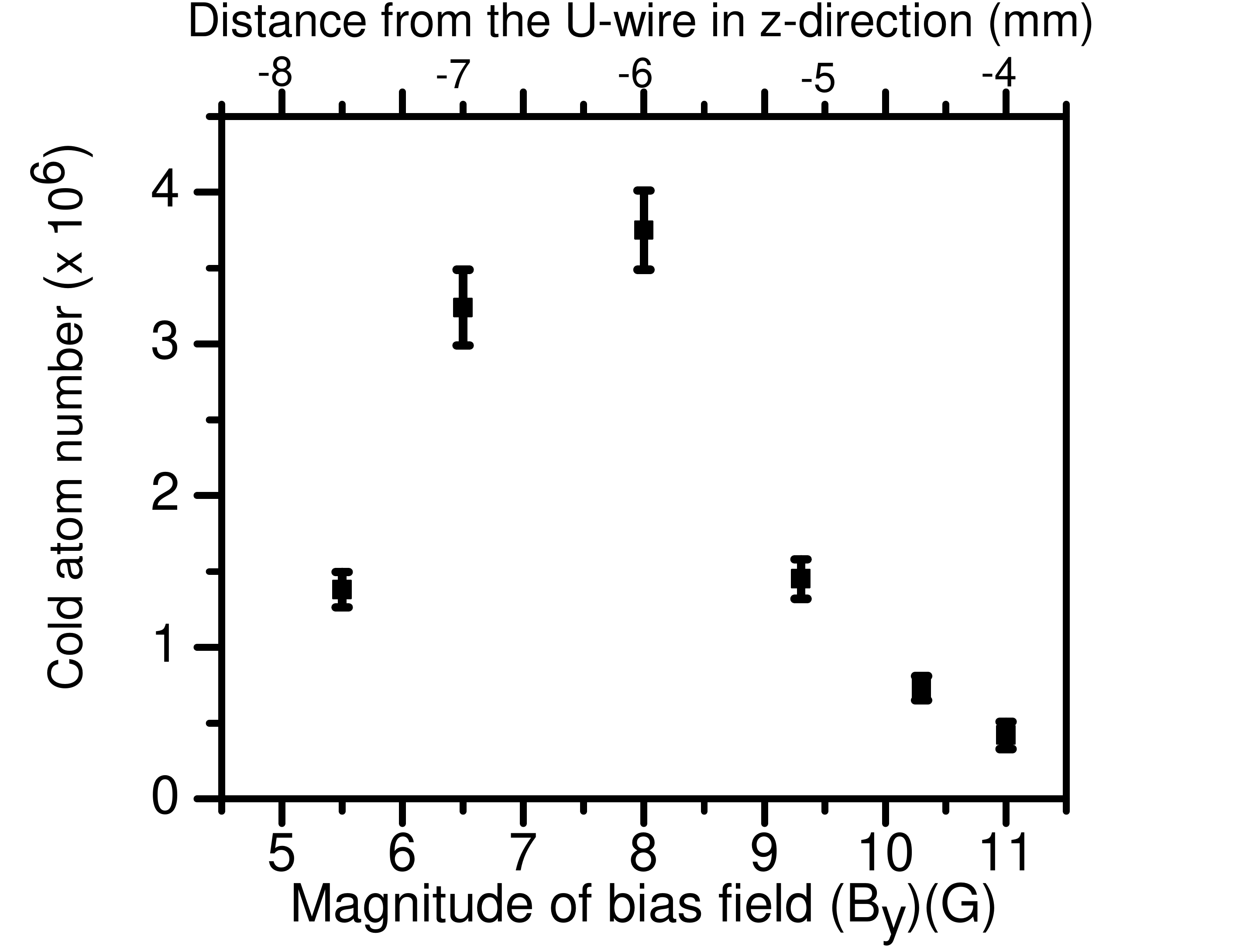}}}
\caption{(a) The images of cold atom cloud showing the shift in the position of MOT cloud along the z-direction ($z$ = - 8 mm to $z$ = - 3 mm ) for different values of bias field ($B_{y}$), with $B_{z}$ = 0. The positions of the atom-chip surface and the U-wire are marked by a horizontal gold and copper color lines, respectively. (b) The variations in the cold atom number with the magnitude of bias field ($B_{y}$), with $B_{z}$ = 0 and with distance from the U-wire in z-direction. The cooling laser beam intensity and detuning are 8.5 $mW/cm^{2}$ and $\sim$ - 14 MHz respectively used in these experiments. The error bar shown in figure represents the statistical variation in the repeated measurements.} \label{sample-figure}
\end{figure}
In our experiments with U-MOT, the variation in cold atom number with quadrupole field minimum position has been investigated. The images in the fig. 8(a) of cold atom cloud in the MOT show the shift in the position of cold atom cloud along the z-direction for different values of bias field ($B_{y}$), with $B_{z}$ = 0. This is due to change in the position of the quadrupole field center as $B_{y}$ is varied at fixed $B_{z}$ = 0. For these measurements, all other parameters like cooling and re-pumping laser beam intensities and detunings etc. were kept unchanged. It is also observed that as distance between MOT cloud and chip surface varies, the number of atoms in the cloud also varies, as shown in fig. 8(b). The maximum number of $\sim 3.7 \times 10^{6}$ was obtained with $B_{y}$ = - 8.0 G which corresponds to the cloud position at $z$ = - 6 mm. This corresponds to optimum gradient and quadrupole field orientation close to ideal one. As discussed previously (fig. 2(a) and 2(b)), for lower bias field $B_{y}$ = - 5.5 G, the field gradient $( 2, 5, 4.5)$ G/cm becomes smaller than that for $B_{y}$ = - 8.0 G. Consequently, with lower bias field $B_{y}$ = - 5.5 G a shallow trap is obtained which results poor number in-spite of large capture volume \cite{yan}. For higher bias field $B_{y}$ = - 11.0 G, as compared to $B_{y}$ = - 8.0 G, though the field gradient $(5, 17, 19)$ G/cm is steeper, smaller capture volume results in lower number of atoms in the MOT. \\
\begin{figure}[ht]
                \centering
                \includegraphics[width=7.5 cm]{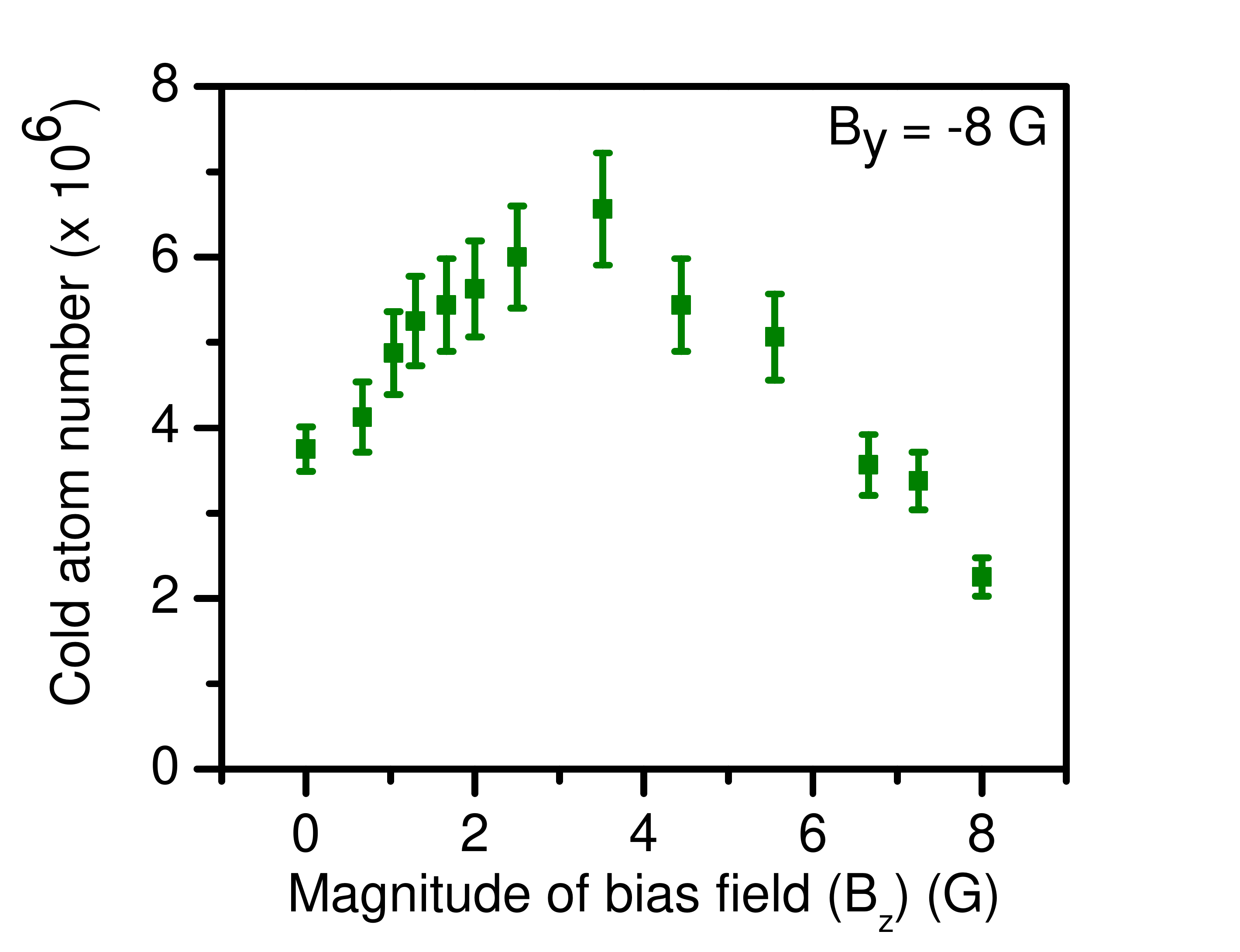}
                \caption{ The variation in number of atoms in the U-MOT with bias magnetic field $B_{z}$, for a fixed value of $B_{y}$ = -8.0 G. The cooling laser beam intensity and detuning are kept at 8.5 $mW/cm^{2}$ and $\sim$ -14 MHz values, respectively, during the experiments. The error bar shown in figure represents the statistical variation in the repeated measurements.}
                \end{figure}
Further, the effect of bias field in z-direction on the MOT formation is also investigated. Figure 9 shows the variation in the number of atoms in the MOT with bias field $B_{z}$, for a fixed value of $B_{y}$ = - 8.0 G. It is found that as we increase the bias field, the number of atoms increases and reaches maximum ($\sim 6.6 \times 10^{6}$) for bias field of $B_{z}$ $\sim$ - 3.5 G. With further increase in $B_{z}$, the number decreases. This is consistent with our results of simulations (fig. 4) that the field configuration with $B_{z}$ $\sim$ - 3.5 G is close to the ideal quadrupole field for MOT. As $B_{z}$ is changed from this optimum value, the resulting field orientation gets deteriorated from the ideal quadrupole field configuration.
\section{Conclusion}
 In conclusion, the loading of MOT with U-wire and applied bias fields has been studied. The magnetic field configuration due to U-wire in presence of a bias field has been simulated. It is shown by simulations that the quadrupole field structure and the center of the quadrupole field vary depending upon the current in U-wire and bias field values. The field configuration, close to ideal quadrupole field configuration, is obtained for a given values of U-wire current and bias magnetic fields values. The observed number in the MOT-cloud and the position of cloud is found to be critically dependent on the U-wire current and bias field values. The observed number of atoms in the MOT is maximum for those current and bias field values which result in the field configuration to be close to the ideal quadrupole field. These studies can be useful in optimisation of the number of atoms in the MOT before transfer and trapping of atoms in the atom-chip wire trap. \\
  \section{Acknowledgement}
We acknowledge Amit Chaudhary for his help during the experiments. We are thankful to Ajay Kak for fabrication of vacuum feed-throughs and R. Shukla and C. Mukherjee for fabrication of atom-chip.



\section{References}
      \begin{thebibliography}{99}
         \bibitem{Folman1}
        Folman, R.; Kruger, P.; Schmiedmayer, J.; Denschlag, j.; Henkel, C. Microscopic atom optics: From wires to an atom chip. \emph{Adv. At. Mol. Opt. Phys.} \textbf{2002}, \emph{48}, 263.
       \bibitem{Folman2}
       Keil, M.; Amit, O.; Zhou, S.; Groswasser, D.; Japha, Y.; Folman, R.  Fifteen years of cold matter on the atom chip: promise, realizations, and prospects. \emph{J. Mod. Opt.} \textbf{2017}, \emph{63}, 1840.
      \bibitem{Dick}
       Dickerson, S.M.; Hogan, J.M.; Sugarbaker, A.; Johnson, D.M.S.; Kasevich, M.A. Multiaxis Inertial Sensing with Long-Time Point Source Atom Interferometry.\emph{Phys.Rev. Lett.} \textbf{2013}, \emph{111}, 083001.
         \bibitem{schumm}
        Schumm, T.; Hofferberth, S.; Anderson, L.M.; Widermuth, S.; Groth, S.; Bar-Joseph, I.; Schmiedmayer, J.; Kruger, P. Matter-wave interferometry in a double well on an atom chip. \emph{Nature Physics} \textbf{2005}, \emph{1}, 57.
       \bibitem{wilder}
       Widermuth, S.; Hofferberth, S.; Lesanovsky, I.; Haller, E.; Anderson, L.M.; Groth, S.; Bar-Joseph, I.; Kruger, P.; Schmiedmayer, J. Microscopic magnetic-field imaging. \emph{Nature} \textbf{2005}, \emph{435}, 440.
       \bibitem{Bohi}
       Bohi, P.; Riedel,M.F.; Hansch, T. W.; Treutlein, P. Imaging of microwave fields using ultracold atoms. \emph{Appl. Phys. Lett.} \textbf{2005}, \emph{435}, 440.
      \bibitem{abend}
       Abend,S.; Gebbe, M.; Gersemann, M.; Ahlers, H.; Muntinga, H.; Giese, E.; Gaaloul, N.;Schubert, C.; Lammerzahl, C.;Ertmer, W.;Schleich, W.P.; Rasel, E.M. Atom-Chip Fountain Gravimeter.  \emph{Phys Rev Lett} \textbf{2016}, \emph{117}, 203003.
      \bibitem{raab}
      Raab, E. L.; Prentis, M.; Cable, A.; Chu, S.; Pritchard, D. E. Trapping of neutral Sodium atoms with Radiation Pressure. \emph{Phys Rev Lett} \textbf{1987}, \emph{59}, 2631.
      \bibitem{clifford}
      Clifford, M. A.; Lancaster, G. P. T.; Mitchell, R. H.; Akerboom, F.; Dholakia, K. Realization of mirror magneto-optical trap. \emph{J. Mod. Opt.} \textbf{2001}, \emph{48}, 1123.
      \bibitem{wilder2}
      Widermuth, S.;Kruger, P.; Becker, C.; Brajdic, M.; Haupt, S.;  Kasper, A.; Folman, R.; Schmiedmayer, J. Optimized magneto-optical trap for experiments with ultracold atoms near surfaces. \emph{Phys Rev A} \textbf{2004}, \emph{69}, 030901(R).
      \bibitem{aubin}
      Aubin, S.; Myrskog, S.; Extavour, M.H.; Leblanc, L. J.; Mckay, D.;Stummer, A.; Thywissen, J. H.  Rapid sympathetic cooling to Fermi degeneracy on a chip. \emph{Nature Physics} \textbf{2006}, \emph{2}, 384.
   
      \bibitem{hansel}
      Hansel, W.; Hommelhoff, P.; Hansch, T. W.; Reichel, J.  Bose-Einstein condensation on a microelectronic chip. \emph{Nature} \textbf{2001}, \emph{413}, 498.
       \bibitem{pkruger}
       Kruger P., PhD thesis.  Coherent matter waves near surfaces, \emph{Ruperto-Carola University of Heidelberg, Germany} \textbf{2004}.
      \bibitem{vivek}
      Singh, Vivek; Tiwari, V. B.; Mishra, S. R.; Rawat, H. S. A tunable Doppler-free dichroic lock for laser frequency stabilization. \emph{Appl. Phys. B} \textbf{2016}, \emph{122}, 225.
      \bibitem{steck}
      Steck, D. A. Rubidium 87 D line data, available online at http:// steck.us/alkalidata.
      \bibitem{yan}
      Hui, Y.; Qing, Y. G.; Jin, W.; Sheng, Z. M. Directly Trapping Atoms in a U-Shaped Magneto-Optical Trap Using a Mini Atom Chip. \emph{Chin. Phys. Lett.} \textbf{2008}, \emph{25}, 3219.
      \end{thebibliography}
\end{document}